\documentclass{svproc}
\usepackage{graphicx}
\usepackage{mathtools}
\usepackage{amsmath,amsfonts,amssymb}
\usepackage{tikz}
\usetikzlibrary {arrows.meta,bending,positioning}
\usepackage{tikz-dimline}
\usepackage{float}
\usepackage{subcaption}
\usepackage{bm}
\usepackage{multirow}
\usepackage{siunitx}
\usepackage{comment}
\usepackage{listings}
\usepackage{hyperref}
\hypersetup{pdftex=true, colorlinks=true, breaklinks=true, linkcolor=blue, menucolor=blue, citecolor=blue, urlcolor=blue}
\usepackage{xcolor}
\definecolor{beaublue}{rgb}{0.94, 0.97, 1.0}
\usepackage{url}

\begin{document}
\mainmatter              
\title{\texttt{PyTOPress}: Python code for topology optimization with design-dependent pressure loads}
\titlerunning{\texttt{PyTOPress}: Python code for topology optimization with  pressure loads}  
%
\author{Shivajay Saxena\inst{1}, Swagatam Islam Sarkar\inst{2} \and
	Prabhat Kumar\inst{2,3,4}}
\authorrunning{S. Saxena, S.I. Sarkar, P. Kumar} 
%

%
\institute{Department of Computer Science and Engineering, \\National Institute of Technology Raipur, Chhattisgarh 492010, India
	\and Department of Mechanical and Aerospace Engineering, \\Indian Institute of Technology Hyderabad, Telangana 502285, India
	\and Department of Computational Engineering, \\Indian Institute of Technology Hyderabad, Telangana 502285, India\\
	\and Department of Engineering Science, \\Indian Institute of Technology Hyderabad, Telangana 502285, India\\
	\email{\url{pkumar@mae.iith.ac.in}}}

\maketitle              

\begin{abstract}
Python is a low-cost and open-source substitute for the MATLAB programming language. This paper presents ``\texttt{PyTOPress}", a compact Python code meant for pedagogical purposes for topology optimization for structures subjected to design-dependent fluidic pressure loads. \texttt{PyTOPress}, based on the ``\texttt{TOPress}" MATLAB code \cite{kumar2023topress}, is built using the \texttt{NumPy} and \texttt{SciPy} libraries. The applied pressure load is modeled using the Darcy law with the conceptualized drainage term. From the obtained pressure field, the constant nodal loads are found. The employed method makes it easier to compute the load sensitivity using the adjoint-variable method at a low cost. The topology optimization problems are solved herein by minimizing the compliance of the structure with a constraint on material volume. The method of moving asymptotes is employed to update the design variables. The effectiveness and success of \texttt{PyTOPress} code are demonstrated by optimizing a few design-dependent pressure loadbearing problems. The code is freely available at \url{https://github.com/PrabhatIn/PyTOPress}.
\end{abstract}
\section{Introduction}\label{sec1}
Nowadays, topology optimization (TO), a computational design technique, is widely used in engineering. The related optimization process extremizes the desired objective under given physical/geometrical constraints \cite{sigmund2013topology}. With the optimization evolution process, the input load on structures may change or remain constant. The former loads are termed design-dependent forces, which are encountered in various applications~\cite{kumar2023topress,kumar2020topology,kumar2022topology}. Modeling such loads within a TO framework is complex and challenging since their position, amount, and direction change as optimization advances. To ease the learning process, Kumar~\cite{kumar2023topress} provides the first publicly available educational MATLAB code to optimize structures subjected to design-dependent pressure loads. Python, an open-source substitute for the MATLAB programming language, is relatively more affordable. However, surprisingly there are only a few published TO programs with Python interfaces \cite{zuo2015simple,smit2021topology,agarwal2023pyhextop,ChadhaPyTOaCNN}. Therefore, an educational code in the Python interface for TO subjected to design-dependent pressure loads is presented herein that can open new avenues for those who use Python but are new to the TO field.

This work introduces ``\texttt{PyTOPress}", a standalone Python code explicitly designed for Python users with limited exposure to TO for pedagogical purposes. Python can be an excellent choice for implementing the TO technique because it is a dynamic and versatile programming language with great utility in many domains. Its powerful libraries and frameworks, such as \texttt{Numpy} and \texttt{SciPy}, are best suited for scientific computation, data analysis, and numerical optimization. \texttt{PyTOPress} is developed per \texttt{TOPress} MATLAB code \cite{kumar2023topress} while using the same nomenclature for the variables and parameters. It provides an easy-to-use platform that encourages further developments and expansions of the field.

The remainder of this paper is arranged as follows. Sec. \ref{sec2} presents the problem formulation. In Sec. \ref{sec3}, we detailed the Python implementations of \texttt{PyTOPress}. In Sec. \ref{sec4}, we explain various extensions of our \texttt{PyTOPress} and their results. Finally, we present the concluding remarks in Sec. \ref{sec6}.

\section{Problem Formulation}\label{sec2}
The idea of using Python, a high level, open-source programming language, for TO is already shown in \cite{zuo2015simple,smit2021topology,agarwal2023pyhextop,ChadhaPyTOaCNN}. Thus, using the language for the design-dependent pressure loading problems, opens up new directions for research and development in TO. Next section provides optimization formulation and sensitivity analysis in brief. A complete description can be found in~\cite{kumar2023topress}.

We showcase three TO problems depicted in Fig. \ref{fig:Pressure_loaded_structures} to demonstrate the code abilities and success. The design domains of an internally pressurized beam, a pressurized piston, and a pressurized chamber are shown in Fig. \ref{fig:Inter_press_arch_Design}, \ref{fig:press_piston}, and \ref{fig:press_chamber}, respectively. All the dimensions and related parameters are adopted from \cite{kumar2023topress}.

\begin{figure}[h!]
	\centering
	\begin{subfigure}{0.3\textwidth}
		\begin{tikzpicture}[scale=0.40]
			\fill [gray!50] (0,0) rectangle (8,4);
			\fill [cyan] (0,-0.3) rectangle (8,0);
			\draw[black, very thick] (0,0) -- (8,0);
			\draw[black, very thick] (8,0) -- (8,4);
			\draw[black, very thick] (8,4) -- (0,4);
			\draw[black, very thick] (0,4) -- (0,0);
			\filldraw[black] (0,0) circle (2.5pt);
			\filldraw[black] (8,0) circle (2.5pt);
			\draw (0,0) -- (-0.4,-0.4) -- (0.4,-0.4) -- cycle;
			\draw (8,0) -- (7.6,-0.4) -- (8.4,-0.4) -- cycle;
			\fill [black] (-0.4,-0.4) rectangle (0.4,-0.6);
			\fill [black] (7.6,-0.4) rectangle (8.4,-0.6);
			\draw[red, very thick, -stealth] (0,-1) -- (0,0);
			\draw[red, very thick, -stealth] (2,-1) -- (2,0);
			\draw[red, very thick, -stealth] (4,-1) -- (4,0);
			\draw[red, very thick, -stealth] (6,-1) -- (6,0);
			\draw[red, very thick, -stealth] (8,-1) -- (8,0);
			\draw[red, very thick, -stealth] (1,-1) -- (1,0);
			\draw[red, very thick, -stealth] (3,-1) -- (3,0);
			\draw[red, very thick, -stealth] (5,-1) -- (5,0);
			\draw[red, very thick, -stealth] (7,-1) -- (7,0);
			\node at (4.4,-1.2) {$p$};
		\end{tikzpicture}
		\caption{Internally pressurized structure}
		\label{fig:Inter_press_arch_Design}
	\end{subfigure}
	\begin{subfigure}{0.3\textwidth}
		\begin{tikzpicture}[scale=0.40]
			\fill [gray!50] (0,0) rectangle (8,4);
			\fill [cyan] (0,4.3) rectangle (8,4);
			\draw[brown!70, very thick] (0,0) -- (8,0);
			\draw[black, very thick] (8,0) -- (8,4);
			\draw[black, very thick] (8,4) -- (0,4);
			\draw[black, very thick] (0,4) -- (0,0);
			\filldraw[black] (4,0) circle (2.5pt);
			\draw (4,0) -- (3.6,-0.4) -- (4.4,-0.4) -- cycle;
			\draw[color=black, very thick](8.1,3.8) circle (0.1);
			\draw[color=black, very thick](8.1,2.9) circle (0.1);
			\draw[color=black, very thick](8.1,2) circle (0.1);
			\draw[color=black, very thick](8.1,1.1) circle (0.1);
			\draw[color=black, very thick](8.1,0.2) circle (0.1);
			\fill [black] (8.2,4) rectangle (8.3,0);
			\fill [black] (3.6,-0.4) rectangle (4.4,-0.6);
			\draw[red, very thick, -stealth] (0,5) -- (0,4);
			\draw[red, very thick, -stealth] (2,5) -- (2,4);
			\draw[red, very thick, -stealth] (4,5) -- (4,4);
			\draw[red, very thick, -stealth] (6,5) -- (6,4);
			\draw[red, very thick, -stealth] (8,5) -- (8,4);
			\draw[red, very thick, -stealth] (1,5) -- (1,4);
			\draw[red, very thick, -stealth] (3,5) -- (3,4);
			\draw[red, very thick, -stealth] (5,5) -- (5,4);
			\draw[red, very thick, -stealth] (7,5) -- (7,4);
			\draw[color=black, very thick](-0.1,3.8) circle (0.1);
			\draw[color=black, very thick](-0.1,2.9) circle (0.1);
			\draw[color=black, very thick](-0.1,2) circle (0.1);
			\draw[color=black, very thick](-0.1,1.1) circle (0.1);
			\draw[color=black, very thick](-0.1,0.2) circle (0.1);
			\fill [black] (-0.2,4) rectangle (-0.3,0);
			\node at (4.4,3.5) {$p=1$};
			\node at (4.6,0.4) {$p=0$};
		\end{tikzpicture}
		\caption{Pressurized piston}
		\label{fig:press_piston}
	\end{subfigure}
	\begin{subfigure}{0.3\textwidth}
		\begin{tikzpicture}[scale=0.45]
			\fill [gray!50] (0,0) rectangle (6,4);
			\fill [cyan] (0,-0.3) rectangle (6,0);
			\draw[black, very thick] (0,0) -- (2.8,0);
			\draw[black, very thick] (3.2,0) -- (6,0);
			\draw[black, very thick] (6,0) -- (6,1.7);
			\draw[black, very thick] (6,2.3) -- (6,4);
			\draw[black, very thick] (6,4) -- (0,4);
			\draw[black, very thick] (0,4) -- (0,0);
			\draw[red, very thick, -stealth] (1.25,-1) -- (1.25,0);
			\draw[red, very thick, -stealth] (2.5,-1) -- (2.5,0);
			\draw[red, very thick, -stealth] (3.75,-1) -- (3.75,0);
			\draw[red, very thick, -stealth] (5,-1) -- (5,0);
			\filldraw[black] (0,0) circle (2.5pt);
			\filldraw[black] (6,0) circle (2.5pt);
			\draw (0,0) -- (-0.4,-0.4) -- (0.4,-0.4) -- cycle;
			\draw (6,0) -- (5.6,-0.4) -- (6.4,-0.4) -- cycle;
			\fill [black] (-0.4,-0.4) rectangle (0.4,-0.6);
			\fill [black] (5.6,-0.4) rectangle (6.4,-0.6);
			\fill [cyan] (42/15,0) rectangle (48/15,23/10);
			\fill [cyan] (48/15,23/10) rectangle (6,17/10);
			\draw[black, very thick] (2.8,0) -- (2.8,2.3);
			\draw[black, very thick] (2.8,2.3) -- (6,2.3);
			\draw[black, very thick] (3.2,0) -- (3.2,1.7);
			\draw[black, very thick] (3.2,1.7) -- (6,1.7);
			\fill [brown] (4,2.3) rectangle (6,2.5);
			\fill [brown] (4,1.5) rectangle (6,1.7);
			\node at (4.4,-0.4) {$p$};
		\end{tikzpicture}
		\caption{Pressurized chamber}
		\label{fig:press_chamber}
	\end{subfigure}
	\caption{Design domain of the (\subref{fig:Inter_press_arch_Design}) internally pressurized beam, (\subref{fig:press_piston}) pressurized piston and (\subref{fig:press_chamber}) pressurized chamber.}  \label{fig:Pressure_loaded_structures}
\end{figure}

\subsection{Optimization Formulation}\label{subsec1}
The compliance minimization problem is solved, for that, one can write the TO formulation as~\cite{kumar2023topress}:

\begin{equation} \label{Eq:OPTI} 
	\begin{rcases}
		\begin{split}
			&{\min_{\tilde{\bm{\rho}}}} \quad C({\tilde{\bm{\rho}}}) = \mathbf{u}^T \mathbf{K}(\tilde{\bm{\rho}})\mathbf{u} = \sum_{j=1}^{\mathtt{nel}}\mathbf{u}_j^T\mathbf{k}_j(\rho_j)\mathbf{u}_j\\
			&\text{subjected to:}\\
			&\bm{\lambda}_1:\,\,\mathbf{A} \mathbf{p} = \mathbf{0}\\
			&\bm{\lambda}_2:\,\,\mathbf{K} \mathbf{u} = \mathbf{F} = -\mathbf{T}\mathbf{p}\\
			&\Lambda\,\,\,:\,V(\tilde{\bm{\rho}})-V^* \le 0\\
			&\quad\,\,\,\,\quad \bm{0} \leq \bm{\rho} \leq \bm{1}
		\end{split}
	\end{rcases},
\end{equation} 

\noindent where \texttt{nel} is the total number of elements used to parameterize the design domain, and \textit{C}($\tilde{\bm{\rho}}$) indicates the structure's compliance. $\mathbf{K(\tilde{\bm{\rho}})}$ and $\mathbf{u}$ represent the global stiffness matrix and displacement vector, respectively. 
$V^*$ and $V$ represent the permitted and current volume of the evolving domain, respectively. Pressure loads give rise to the global force vector $\mathbf{F}$, and $\mathbf{T}$ is the global transformation matrix. $\mathbf{p}$ and $\mathbf{A}$ are the global pressure vector and flow matrix, respectively. $\bm{\lambda}_1$ (vector), $\bm{\lambda}_2$ (vector), and $\Lambda$ (scalar) are the Lagrange multipliers. The filtered design vector $\tilde{\bm{\rho}}$ corresponds to the design variable vector $\bm{\rho}$ which is determined using the classical density filter \cite{bruns2001}.

The filtering is prepared using the \texttt{scipy.ndimage.correlate} method in the Python code. The state equation $\mathbf{K} \mathbf{u} = \mathbf{F} = -\mathbf{T}\mathbf{p}$, where $\mathbf{p}$, the pressure field, is obtained by solving $\mathbf{A} \mathbf{p} = \mathbf{0}$. 
The modified Solid Isotropic Material with Penalization (SIMP) approach is used to interpolate Young's modulus of element $i$ as :

\begin{equation}
	\mathrm{E}_i = \mathrm{E}_\mathrm{min} + \tilde{\rho}^p_i(\mathrm{E}_0-\mathrm{E}_\mathrm{min}),
\end{equation}

\noindent where the penalization factor $p$, encourages TO to converge towards 0-1 solutions, $p=3$ is used in this paper. $\mathrm{E}_0$ is the elemental stiffness with $\tilde{\rho}_i=1$, $\mathrm{E}_\mathrm{min}$ denotes a minimal  stiffness assigned to void regions to prevent the stiffness matrix from becoming singular.

\subsection{Sensitivity analysis}\label{subsec2}
The MMA \cite{svanberg1987method} is used to update the design variables; thus, the objective's and constraints' derivatives are needed. We determine the sensitivities by applying the adjoint-variable method. One can refer to Kumar~\cite{kumar2023topress} for a detailed description. The sensitivity is written as: 

\begin{equation}\label{Eq:senstivities_Obj}
	\frac{d {C}}{d \tilde{\bm{\rho}}} = -\mathbf{u}^T \frac{\partial\mathbf{K}}{\partial \tilde{\bm{\rho}}}\mathbf{u} + \underbrace{2\mathbf{u}^T \mathbf{T} \mathbf{A}^{-1}\frac{\partial\mathbf{A}}{\partial\tilde{\bm{\rho}}}\mathbf{p}}_{\text{Load sensitivities}}.
\end{equation}

One notes that the load sensitivities change the compliance sensitivities (Eq.~\ref{Eq:senstivities_Obj}). With a constant load, the compliance sensitivities contain only the first term of Eq.~\ref{Eq:senstivities_Obj}. The load sensitivities should not be neglected as they can affect the optimized shape, topology, and performances~\cite{kumar2020topology,kumar2023topress}. The chain rule is  used to find the derivatives of the objective function with respect to the design variables~\cite{kumar2023topress}. It is easy to find the derivative of the volume constraint~\cite{kumar2023honeytop90}. Next, we describe the Python code line-by-line.

\section{Python Implementation}\label{sec3}

This section provides an implementation of PyTOPress code, \texttt{PyTOPress}.

\subsection{Import Libraries: (lines 1 to 12)}\label{subsec3}

First, required libraries are imported and classified into core and specialized ones. Additionally, the MMA module is customized to update the design variable.

\subsubsection{Core Libraries}\label{subsubsec1}

\begin{itemize}
	\item\textbf{scipy}: A comprehensive library for scientific computing, providing functions for optimization, linear algebra, integration, interpolation, etc.
	\item\textbf{numpy (as np)}: It handles large, multi-dimensional arrays and matrices. It also includes many advanced mathematical functions.
	\item\textbf{matplotlib.pyplot (as plt)}: It creates visualizations, including plots and images.
\end{itemize}

\subsubsection{Specialized Libraries}\label{subsubsec2}
\begin{itemize}
	\item\textbf{numpy.matlib}: It is specifically designed for matrix operations, offering functions for creating and manipulating matrices.
	\item\textbf{scipy.ndimage}: It contains image processing tools, such as this code's filtering operations.
	\item\textbf{scipy.sparse}: It efficiently handles sparse matrices, which is crucial for large-scale problems.
	\item\textbf{scipy.sparse.linalg}: It provides solvers for sparse linear systems, essential for solving equations involving sparse matrices.
\end{itemize}

\subsubsection{Custom Functions and Modules}\label{subsubsec3}
\begin{itemize}
	\item\textbf{mmasub}: A customized version of the MMA optimization algorithm to solve the specific problem.
\end{itemize}

The code effectively solves the TO problems with the combination of these libraries.

\subsection{\texttt{PyTOPress} Function Definition: (line 14)}\label{subsec4}

\texttt{PyTOPress} function is defined as follows :
\begin{lstlisting}[basicstyle=\scriptsize\ttfamily,breaklines=false,numbers=none,frame=tb,backgroundcolor=\color{beaublue},language=Python]
	def PyTOPress(nelx, nely, volfrac, penal, rmin, etaf, betaf, lst, maxit)
\end{lstlisting}
where, \texttt{nelx} and \texttt{nely} define the dimensions of the finite element mesh, specifying the number of elements in the $x$ and $y$ directions, respectively. \texttt{volfrac} represents the desired volume fraction. \texttt{penal} is the penalization factor used to penalize intermediate density values. \texttt{rmin} describes the minimum filter radius that controls the size of the filtering region. \texttt{etaf} and \texttt{betaf} are flow parameters~\cite{kumar2023topress}. The involvement of load sensitivities is determined by \texttt{lst}; i.e., \texttt{lst} = 1 denotes that load sensitivities are included, whereas \texttt{lst} = 0 denotes the opposite. \texttt{maxit} sets the maximum number of optimization iterations.

\subsection{Part 1: Material and Flow Parameters: (lines 15 to 20)}\label{subsec5}

Young's modulus of the material, that of the void regions, and Poisson's ratio of the material are stored in \texttt{E1} (line 16), \texttt{Emin} (line 17), and \texttt{nu} (line 18), respectively. The values for the flow coefficient of a void element \texttt{Kv}, flow contrast \texttt{epsf}, filter radius \texttt{r}, and penetration depth \texttt{Dels} are assigned on line 19. It also defines drainage parameters \texttt{Ds} and \texttt{kvs} (line 20).

\subsection{Part 2: Finite Element Analysis (FEA) Preparation: (lines 21 to 53)}\label{subsec6}

The total number of elements and nodes are stored in \texttt{nel} and \texttt{nno} (line 22), respectively. We construct a nodal connectivity matrix, referred to as \texttt{nodenrs} (line 23), to establish the relationship between elements and their corresponding nodes. The degrees of freedom (DOFs) for displacement components of each element are defined and stored in \texttt{Udofs} (line 27). Boundary nodes are identified and categorized by their location. \texttt{Lnode}, \texttt{Rnode}, \texttt{Bnode}, and \texttt{Tnode} represent the nodes of the left, right, bottom, and top edges of the structure. Elemental pressure-related DOFs are assigned to \texttt{Pdofs} (line 30). \texttt{allPdofs} and \texttt{allUdofs} represent pressure and displacement DOFs of the discretized domain, respectively (line 30). \texttt{\{iP,jP\}} (lines 31-32), \texttt{\{iT,jT\}} (lines 33-34) and \texttt{\{iK,jK\}} (lines 35-36) contain the rows and columns indices for the flow, transformation, and stiffness matrices, respectively are generated based on the defined DOFs~\cite{kumar2023topress}. Element Darcy flow matrix \texttt{Kp} (line 37), element drainage matrix \texttt{KDp} (line 38), and element transformation matrix \texttt{Te} (line 39-40) are recorded. The elemental stiffness matrix \texttt{ke} is computed on line 45. \texttt{Ts} vector is evaluated by converting \texttt{Te} into a column vector and appropriately reshaping (line 46), which helps to find the global transformation matrix \texttt{TG} (line 47).

\subsection{Part 3: Pressure and Structure Boundary Conditions and Loads: (lines 54 to 63)}\label{subsec7}

The code initializes a minimal pressure value, \texttt{PF}, at all nodal points on line 55, with specific adjustments made for nodes located on the top, bottom, left, and right boundaries. The code differentiates between pressure DOFs that are fixed, \texttt{fixedPdofs} (line 58), and those that are free to vary, \texttt{freePdofs} (line 59). A similar classification is carried out for displacement degrees of freedom, resulting in \texttt{fixedUdofs} (line 61) and \texttt{freeUdofs} (line 62). To begin the computational process, the displacement vector \texttt{U} and Lagrange multipliers \texttt{lam1} are initialized (line 63) with starting values.

\subsection{Part 4: Filter Preparation: (lines 64 to 67)}\label{subsec8}

The weight factor matrix for filtering, \texttt{h}, is constructed (line 66) using a distance-based calculation involving \texttt{np.sqrt(dx**2 + dy**2)} within a specified radius, \texttt{rmin}. \texttt{h} is then subjected to a convolution operation with a unit square represented by \texttt{np.ones((nely, nelx))} using the \texttt{scipy.ndimage.correlate} function. \texttt{Hs} denotes the normalization constant vector responsible for filtering operation (line 67).

\subsection{Part 5: MMA Optimization Preparation and Initialization: (lines 68 to 78)}\label{subsec9}

The derivative of the volume constraint, \texttt{dVol0}, is initialized on line 69 to ensure a uniform initial distribution. The \texttt{x[act - 1]} array is used to assign initial densities to active elements (line 70), excluding any potential non-design regions. The number of design variables and constraints, indicated by \texttt{nMMA} and \texttt{mMMA}, respectively, are defined on line 71, with \texttt{mMMA} set to 1 for the volume constraint. Copies of the design variables \texttt{xphys} for filtering and \texttt{xMMA} for the MMA algorithm are created on line 71. The move limit for MMA is specified in \texttt{mvLt} (line 71). Lower and upper bounds for the design variables are defined on line 72 as \texttt{xminvec} and \texttt{xmaxvec}, respectively, with corresponding copies \texttt{low} and \texttt{upp} defined on line 73. The MMA constants are set on line 74, with \texttt{cMMA} defined as 1000 and \texttt{dMMA} as 0, \texttt{a0} is set to 1, and \texttt{aMMA} to 0. To track changes in the design variables, \texttt{xold1} and \texttt{xold2} store values from the previous iteration, as initialized on line 75. Counters \texttt{loop} and \texttt{change} are initialized (line 76) to monitor the optimization iterations and design variable changes. Finally, the volume sensitivity is calculated and assigned to \texttt{dVol} (line 77).

\subsection{Part 6: MMA Optimization Loop: (lines 79 to 139)}\label{subsec10}
The optimization loop starts at line 80 as follows :
\begin{lstlisting}[basicstyle=\scriptsize\ttfamily,breaklines=false,numbers=none,frame=tb,backgroundcolor=\color{beaublue},language=Python]
	while loop < maxit and change > 0.01:
	loop += 1
\end{lstlisting}
This loop iterates for a maximum of \texttt{maxit} iterations or until the change in design variables falls below a threshold (0.01). Each iteration involves several steps as follows:\\
\noindent \textbf{Part 6.1: Solving Flow Balance Equation:}\\
The flow coefficient, \texttt{Kc} (line 83), and drainage term, \texttt{Dc} (line 84), are calculated based on the current design variables, \texttt{xphys}, and flow parameters \texttt{etaf} and \texttt{betaf}. On line 85, the sparse matrix \texttt{Ae}, representing the flow vector, is constructed using predefined element matrices \texttt{Kp}, \texttt{KDp}, and \texttt{Hs}. The global flow matrix, \texttt{AG}, is assembled on line 86 using \texttt{Ae} and sparse matrix operations provided by \texttt{scipy.sparse.csr\_matrix}. The pressure values at the free DOFs, \texttt{PF[freePdofs - 1]}, are computed on line 92 by solving a sparse linear system using the \texttt{spsolve} function.

\noindent \textbf{Part 6.2: Determining Nodal Loads and Global Displacement Vector:}\\
The global stiffness matrix \texttt{KG} is prepared on line 98 using sparse matrix operations. The global displacements at free degrees of freedom, \texttt{U[freeUdofs-1]}, are determined (line 103) by solving a sparse linear system. The right-hand side of this system is the force vector \texttt{F}, which is assembled (line 95) from pressure and the global transformation matrix \texttt{TG}.

\noindent \textbf{Part 6.3: Objective, Constraint, and Sensitivities Computation:}\\
The compliance of the structure, representing the objective function \texttt{obj}, is computed on line 105. The vector \texttt{lam1}, which indicates the Lagrange multiplier, is determined on line 106. The vector \texttt{objsT1}, containing the first term of the right-hand side of Eq.~\ref{Eq:senstivities_Obj}, is calculated on line 107 based on the element stiffness matrix \texttt{ke}, displacements \texttt{U}, and the penalization factor \texttt{penal}. The load sensitivity, \texttt{objsT2}, is computed on line 110, taking into account \texttt{Hs}, \texttt{kvs}, \texttt{Ds}, pressure, and Lagrange multipliers. These two terms are combined using \texttt{lst} to form the overall sensitivity of the objective function, \texttt{objsens}, on line 111. Finally, the sensitivity is normalized (lines 113-115) by dividing it by a pre-computed value, \texttt{normf}.

\noindent \textbf{Part 6.4: Setting and Calling MMA Optimization:}\\
The variable bounds, \texttt{xminvec} and \texttt{xmaxvec}, are adjusted on lines 118-119 based on the current design variables, \texttt{xval}, and the permitted movement limit, \texttt{mvLt}. The \texttt{mmasub} function, a custom implementation of the MMA optimization algorithm, is then invoked to update the design variables, \texttt{xmma}. This function utilizes the normalized objective function, \texttt{(obj * normf)}, the sensitivities, \texttt{objsens}, the total volume, \texttt{Vol}, and its sensitivity, \texttt{dVol}, as input parameters. To maintain a history of changes in the design variables, the variables \texttt{xold1} and \texttt{xold2} are updated on line 122. Finally, the overall change in the design variables is calculated and stored in the \texttt{change} variable on line 123.

\noindent \textbf{Part 6.5: Printing and Plotting Results:}\\
The code iterates through optimization steps, performing calculations and updates within each loop. Below is a breakdown of the final step:

\textbf{Printing Results:}\ The variables \texttt{loop}, \texttt{obj * normf}, \texttt{np.mean(xphys)}, and \texttt{change} represents the current iteration number, the normalized objective function value, the average density of the design, and the magnitude of change in the design variables, respectively are printed and displayed on line 131.

\textbf{Density Image Visualization:}\ The design variables, \texttt{xphys}, are reshaped into a 2D array to represent an image on line 133. Depending on the implementation, the reshaped array may be inverted for void and solid representation. Using \texttt{plt.imshow}, the density image is displayed with a grayscale colormap \texttt{cmap=`gray'} on line 135. The image display limits are set to a minimum density of 0 and a maximum density of 1. The plot is shown using \texttt{plt.show(block=False)}, allowing continuous updates without blocking the code execution on line 138. A fraction of a second pause, \texttt{plt.pause(0.001)}, is introduced to create an animation-like effect as the density image changes with each iteration on line 139.

\subsection{The internally pressurized arch structure (function call, line 140-141)}\label{subsec11}

The function is called (line 141) to design pressurized arch structure (Fig.~\ref{fig:Inter_press_arch_Design}) as :
\begin{lstlisting}[basicstyle=\scriptsize\ttfamily,breaklines=false,numbers=none,frame=tb,backgroundcolor=\color{beaublue},language=Python]
	PyTOPress(200, 100, 0.30, 3, 2.4, 0.2, 8, 1, 100)
\end{lstlisting}
which sets up a $200 \times 100$ mesh, a volume fraction of 0.3, a penalization factor of 3, and a filter radius 2.4. The flow parameters are specified as $\eta_f=0.2$ and $\beta_f=8$. The parameter \texttt{lst} is set to 1, indicating that load sensitivities are included, and the maximum number of iterations is set to 100.\\

\begin{figure}[h!]
	\centering
	\begin{subfigure}{0.3\textwidth}
		\includegraphics[scale=0.25]{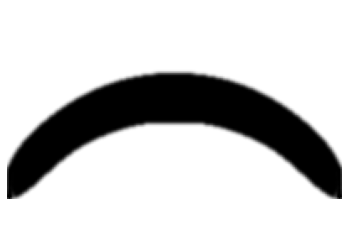}
		\caption{Pressurized arch}
		\label{fig:TOPress_arch}
	\end{subfigure}
	\begin{subfigure}{0.3\textwidth}
		\includegraphics[scale=0.25]{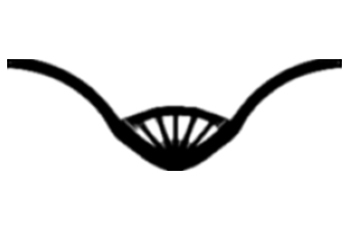}
		\caption{Pressurized piston}
		\label{fig:TOPress_piston}
	\end{subfigure}
	\begin{subfigure}{0.3\textwidth}
		\includegraphics[scale=0.25]{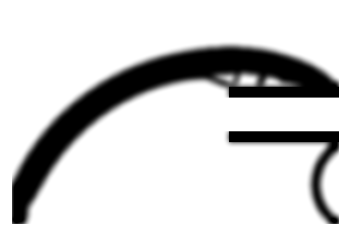}
		\caption{Pressurized chamber}
		\label{fig:TOPress_chamber}
	\end{subfigure}
	\caption{The optmized designs of (\subref{fig:TOPress_arch}) Pressurized arch, (\subref{fig:TOPress_piston}) Pressurized piston, and (\subref{fig:TOPress_chamber}) Pressurized chamber.}  \label{fig:Pressurized_structures}
\end{figure}

The optimized result is shown in Fig. \ref{fig:TOPress_arch}. The result is similar to the one obtained in \cite{kumar2023topress}. The final design looks identical to an arch.

\section{Extension to PyTOPress and optimized results}\label{sec4}

The  extensions of \texttt{PyTOPress} for two problems are presented herein.

\subsection{Pressurized Piston}\label{subsec14}

The first extension is for designing pressurized piston. The design domain, pressure, and boundary conditions for the problem are adopted from \cite{kumar2023topress} and depicted in Fig.~\ref{fig:Pressure_loaded_structures}b. To get the optimized design of the pressurized piston problem first reported in \cite{bourdin2003design}, we modify \texttt{PyTOPress} code in lines 56, 57, and 61 as follows:
\begin{lstlisting}[basicstyle=\scriptsize\ttfamily,breaklines=true,numbers=none,frame=tb,backgroundcolor=\color{beaublue},language=Python]
	PF[Bnode - 1] = 0
	PF[Tnode - 1] = Pin
	fixedUdofs = np.hstack([2*Bnode[math.ceil((nelx+1)/2)]-1, 2*Bnode[math.ceil((nelx+1)/2)],  2*Lnode-1, 2*Rnode-1])
\end{lstlisting}
To obtain the optimized result for this problem, we modify the function call on line 141 of \texttt{PyTOPress} as follows:
\begin{lstlisting}[basicstyle=\scriptsize\ttfamily,breaklines=true,numbers=none,frame=tb,backgroundcolor=\color{beaublue},language=Python]
	PyTOPress(300, 100, 0.20, 3, 2.4, 0.1, 8, 1, 150)
\end{lstlisting}
With these modifications, we obtained the optimized design for the piston problem. The result is shown in Fig. \ref{fig:TOPress_piston} which resembles the one presented in~\cite{kumar2023topress}.

\subsection{Pressurized chamber design}\label{subsec15}

A Pressurized chamber is designed next. The design and non-design domain, pressure, and boundary conditions for the pressurized chamber problem are adopted from \cite{kumar2023topress} (See Fig.~\ref{fig:Pressure_loaded_structures}c). To optimize the design domain, we modify the \texttt{PyTOPress} code by replacing lines 52 to 53 with the following code:
\begin{lstlisting}[basicstyle=\scriptsize\ttfamily,breaklines=true,numbers=none,frame=tb,backgroundcolor=\color{beaublue},language=Python]
	s1 = elNrs[(3 * nely // 8) - 1 : 17 * nely // 40, (2 * nelx // 3) - 1 : nelx]
	s2 = elNrs[(23 * nely // 40) - 1 : 5*nely//8, (2 * nelx // 3) - 1 : nelx]
	v1 = elNrs[(17 * nely // 40) - 1 :, (7 * nelx // 15) - 1 : 8 * nelx // 15]
	v2 = elNrs[(17 * nely // 40) - 1 : 23 * nely // 40, (8 * nelx // 15) - 1 : nelx]
	[NDS, NDV] = np.hstack([s1.flatten(order='F'), s2.flatten(order='F')]), np.hstack([v1.flatten(order='F'), v2.flatten(order='F')])
	act = np.setdiff1d(np.arange(1, nel + 1), np.union1d(NDS, NDV))
	s1fix = elNrs[(3 * nely // 8) - 1 : 17 * nely // 40, nelx - 1]
	s2fix = elNrs[(23 * nely // 40) - 1 : 25 * nely // 40, nelx - 1]
	fixx = np.unique(Pdofs[(np.hstack([s1fix.flatten(order='F'), s2fix.flatten(order='F')])) - 1, :])
	PF[np.hstack([Tnode - 1, Lnode - 1, Rnode - 1])] = 0
	PF[np.hstack([np.unique(Pdofs[NDV - 1, :]) - 1, Bnode - 1])] = Pin
	fixedUdofs = np.hstack([(2*Bnode[0]-1), (2 * Bnode[0]), (2 * Bnode[-1] - 1), (2 * Bnode[-1]), (2 * fixx - 1), (2 * fixx)])
\end{lstlisting}
To obtain the optimized result for this problem, we modify the function call on line 141 of \texttt{PyTOPress} as follows:
\begin{lstlisting}[basicstyle=\scriptsize\ttfamily,breaklines=true,numbers=none,frame=tb,backgroundcolor=\color{beaublue},language=Python]
	PyTOPress(300, 200, 0.20, 3, 6, 0.1, 10, 1, 200)
\end{lstlisting}
With these modifications, we get the optimized pressurized chamber problem results. The result is shown in Fig. \ref{fig:TOPress_chamber}, which resembles the corresponding obtained result in~\cite{kumar2023topress}.

\section{Conclusion}\label{sec6}
In this work, we use the power of open-source libraries to convert the original MATLAB implementation, \texttt{TOPress}~\cite{kumar2023topress} into a Python code, \texttt{PyTOPress}. This conversion unlocks multiple benefits, contributing considerably to accessibility, maintainability, and extensions. The Python code removes licensing obstacles and is easily accessible to a vast research community by taking advantage of Python's open-source nature. The conversion makes use of scientific libraries like \texttt{SciPy} and \texttt{NumPy} as well as the extensive Python ecosystem. The focus is on making the code clear and concise, which improves its readability and maintainability. The original state of \texttt{PyTOPress} is set for designing pressure loadbearing lid structure. Code's extensions are also provided to solve loadbearing structures with different boundary conditions. Python's simple syntax helps the future development of the code in the field of TO. This opens up new possibilities for investigations and extensions in the Python environment towards 3D problems with pressure loading~\cite{kumar2021topology,kumar2024TOPress3D}.


\begin{thebibliography}{12}
	\bibitem{kumar2023topress}
	Kumar, P., 2023. TOPress: a MATLAB implementation for topology optimization of structures subjected to design-dependent pressure loads. Structural and Multidisciplinary Optimization, 66(4), p.97.
	
	\bibitem{sigmund2013topology}
	Sigmund, O. and Maute, K., 2013. Topology optimization approaches: A comparative review. Structural and multidisciplinary optimization, 48(6), pp.1031-1055.
	
	\bibitem{kumar2020topology}
	Kumar, P., Frouws, J.S. and Langelaar, M., 2020. Topology optimization of fluidic pressure-loaded structures and compliant mechanisms using the Darcy method. Structural and Multidisciplinary Optimization, 61, pp.1637-1655.
	
	\bibitem{kumar2022topology}
	Kumar, P., 2022. Topology optimization of stiff structures under self-weight for given volume using a smooth Heaviside function. Structural and Multidisciplinary Optimization, 65(4), p.128.
	
	\bibitem{zuo2015simple}
	Zuo, Z.H. and Xie, Y.M., 2015. A simple and compact Python code for complex 3D topology optimization. Advances in Engineering Software, 85, pp.1-11.
	
	\bibitem{smit2021topology}
	Smit, T., Aage, N., Ferguson, S.J. and Helgason, B., 2021. Topology optimization using PETSc: a Python wrapper and extended functionality. Structural and Multidisciplinary Optimization, 64, pp.4343-4353.
	
	\bibitem{agarwal2023pyhextop}		
	Agarwal, A., Saxena, A. and Kumar, P., 2023. PyHexTop: a compact Python code for topology optimization using hexagonal elements. arXiv preprint arXiv:2310.01968.
	\bibitem{ChadhaPyTOaCNN}	
	Chadha, K.S., Kumar, P., 2023. PyTOaCNN: Topology optimization using an adaptive convolutional neural network in Python. arXiv preprint arXiv:2404.12244
	\bibitem{bruns2001}
	Bruns, T.E. and Tortorelli, D.A., 2001. Topology optimization of non-linear elastic structures and compliant mechanisms. Computer methods in applied mechanics and engineering, 190(26-27), pp.3443-3459.
	\bibitem{svanberg1987method}
	Svanberg, K., 1987. The method of moving asymptotes-a new method for structural optimization. International journal for numerical methods in engineering, 24(2), pp.359-373.
	
	\bibitem{kumar2023honeytop90}
	Kumar, P., 2023. HoneyTop90: A 90-line MATLAB code for topology optimization using honeycomb tessellation. Optimization and Engineering, 24(2), pp.1433-1460.
	
	\bibitem{bourdin2003design}
	Bourdin, B. and Chambolle, A., 2003. Design-dependent loads in topology optimization. ESAIM: Control, Optimisation and Calculus of Variations, 9, pp.19-48.
	
	\bibitem{kumar2021topology}
	Kumar, P. and Langelaar, M., 2021. On topology optimization of design‐dependent pressure‐loaded three‐dimensional structures and compliant mechanisms. International Journal for Numerical Methods in Engineering, 122(9), pp.2205-2220.
	\bibitem{kumar2024TOPress3D}
	Kumar, P., 2024. TOPress3D: 3D topology optimization with design-dependent pressure loads in MATLAB. Optimization and Engineering (2024).
\end{thebibliography}
\end{document}